\documentclass[12pt]{article}

\advance\hoffset by -1.1truecm
\advance\voffset by -1.0truecm

\def\be{\begin{equation}}
\def\ee{\end{equation}}
\def\ba{\begin{eqnarray}}
\def\ea{\end{eqnarray}}

\newcommand{\R}{\mbox{I \hspace{-0.82em} R}}
\newcommand{\x}{{\bf x}}
\newcommand{\p}{{\bf p}}

\def\be{\begin{equation}}
\def\ee{\end{equation}}
\def\bea{\begin{eqnarray}}
\def\eea{\end{eqnarray}}

\begin{document}
\title{Recent results on UV-regularisation through UV-modified uncertainty relations\thanks{Talk presented at the 5th 
Wigner Symposium, Vienna, August 1997}}
\author{Achim Kempf\thanks{Research Fellow of 
Corpus Christi College in the University of Cambridge} \\
Department of Applied Mathematics \& Theoretical Physics\\
Cambridge CB3 9EW, U.K} 

\date{}
\maketitle


\maketitle
\vskip-7.5truecm

\hskip11.7truecm
{\tt DAMTP-97-136} 

\hskip11.7truecm
{\tt hep-th/9711204}
\vskip6.5truecm

{\flushleft{\large {\bf 1~~ Introduction}}}
\medskip\newline
It is well known that when trying to resolve sufficiently
small distances the required test particles of short wave length
and high energy will have a gravitational effect
that significantly disturbs the space
time structure which is being probed. Even simple 
dimensional analysis shows that therefore
a theory of quantum gravity is needed 
the latest at the Planck scale of about $10^{-35}m$. 
The very notions of
points and distances are in question at this scale.
This problem has been approached from various
directions, among the vast
literature, see e.g. \cite{townsend}-\cite{f}.

Whatever final form a fundamental theory 
of quantum gravity at the Planck scale
will take, it seems reasonable to assume that it will express 
spatial information 
through some formal  matrices or operators  $\x_i$.
After all, formal position operators are present
in quantum field theory: technically,
the fields in the path integral
form a representation of a Heisenberg algebra
of hermitean positions $\x_\mu=\x_\mu^*$ and momenta $\p_\mu=\p_\mu^*$ 
which obey the quantum mechanical
commutation relations $[\x_\mu,\p_\nu]=i\hbar \eta_{\mu\nu}$,
see e.g. \cite{dewitt,ak-jmp-reg}. Of course,
these operators do not have a simple quantum mechanical interpretation.
But, their presence in quantum field theory means that 
even if, hypothetically, the fundamental quantum gravity theory 
were not to contain anything close to formal 
position operators, formal linear position operators 
would still have to emerge
{}from the fundamental theory the latest at that scale at which 
$(3+1)$- dimensional quantum field theory becomes a valid description, 
which is probably not very far from the Planck scale.
In the absence of direct experimental access to these scales
it is instructive to analyse the full 
set of possible short distance structures
which linear position operators can describe.
 
To be precise, we make the weak assumption that these $\x_i$ are
elements of an associative, complex and possibly noncommutative 
algebra with the involution in the 
algebra acting as $\x_i^* = \x_i$. Without further assumptions,
what can be said about the set of all possible short distance
structures which can be described in this way? 
{\flushleft{\large {\bf 2~~ 
Self-adjoint versus symmetric position operators} }}
\medskip\newline
Under the above assumption, 
simple functional analytical arguments show that
there are in fact only three distinct short distance structures which can
arise, with all other cases being mixtures of the three:

Consider an element $\x$ in an associative complex algebra ${\cal A}$ 
with the involution acting as $\x^*=\x$. A representation of ${\cal A}$,
and thus of $\x$, on a maximal dense domain $D$ in a Hilbert space $H$ can now 
reveal three possible features of $\x$: The operator
 $\x$ may be self-adjoint, in which
case there are the two possibilities of $\x$ having (I) a discrete spectrum
or (II) a continuous spectrum. Or, and this is the 
third case, (III), $\x$ may not be self-adjoint but merely symmetric.
It will be crucial that in the case III, $\x$ cannot be diagonalised
in $D$.

In the cases I and II, $\x$ describes 
the familiar
 short distance structures of the lattice and the continuum
(or mixtures of both). 
The case III is the only alternative, because any Hilbert space 
representation of the
elements of the algebra together with the involution in the algebra 
ensures at least 
that $\forall ~\vert\psi\rangle \in D:~~\langle \psi\vert \x\vert
\psi\rangle \in \R$, which is to say that the operator $\x$ is symmetric.
As will be explained below, what makes this third short distance structure
an interesting candidate is that it 
does not require the breaking of translation invariance, as opposed
to the lattice, while it indeed also provides an ultraviolet cutoff, as
opposed to the continuum.

{\flushleft{\large {\bf 3~~ 
Uncertainty relations and commutation relations
} }}
\medskip\newline
We recall that symmetry of an operator means that in its domain
all expectation values are real - without however ensuring that it 
is diagonalisable or that it has 
any eigenvectors at all. It will be convenient to consider now 
the uncertainty in positions $\Delta x_{\vert\psi\rangle} =
\langle\psi\vert(\x - \langle\psi\vert\x\vert\psi\rangle)^2\vert\psi
\rangle^{1/2}$ (for $\vert\psi\rangle$ normalised).
 It is well defined on $D$ and we can use it as a (nonlinear)
`indicator functional' $\Delta x: D \rightarrow \R$ because it vanishes
for eigenvectors, and only for eigenvectors.

In the self-adjoint cases I and II clearly $\inf_{\vert \psi\rangle
\in D}\Delta x_{\vert \psi \rangle}=0$.
When $\x$ is merely symmetric this may also hold, as $\x$ may still have
some (normalisable or nonnormalisable) eigenvectors -
we can always tensor a symmetric $\x$ with a self-adjoint
matrix or operator which then contributes its eigenvectors. This would be
a mixed case. 

Let us focus instead on  `pure'  cases of type III, where 
there exists a $\Delta x_0$ such that
$\inf_{\vert \psi\rangle \in D}\Delta x_{\vert \psi \rangle}
= \Delta x_0 >0$. One then deals with
a geometry in which all physical states describe particles which are not
better localised (or not `smaller')
than some finite value $\Delta x_0$. We therefore conclude
that in these cases of type III,
the uncertainty relations must contain correction terms which 
modify their behaviour in the ultraviolet to the effect
 that the smallest position uncertainty is finite.

As a simple case in one dimension, an uncertainty relation with
suitable correction terms takes the form $\Delta x \Delta p
\ge \hbar/2(1 +\beta (\Delta p)^2 + ... )$ which
implies $\Delta x_0 =\hbar \sqrt{\beta}$, 
see e.g.{ }\cite{grossmende,witten}. (Note that not all
choices of correction terms would lead to a case III situation). 
{}From $\Delta A \Delta B \ge 1/2 \vert \langle [A,B] \rangle\vert$
 which holds
for all symmetric $A,B$ we deduce the commutation relations $[\x,\p] =
i\hbar (1 + \beta \p^2 +...)$. The representation theory then indeed shows
that $\x$ is an example of case III, i.e. $\x$ is 
symmetric and non-self-adjoint in all
representations of the algebra ${\cal A}$ generated by $\x$ and $\p$,
 as first shown in { }\cite{ucr}.

{\flushleft{\large {\bf 4~~ 
External symmetry and UV regularisation
} }}
\medskip\newline
Concerning the preservation of symmetries, 
we can  read off from the above commutation
relation that it is invariant under translations $\x \rightarrow \x+a$,
$\p\rightarrow \p$. Indeed, as shown in { }\cite{osc}, 
in arbitrary dimensions both translation and 
rotation invariance can be preserved. The first order correction terms under
this constraint are unique. The case III short distance structure is 
as well compatible with
generalised symmetries, such as the quantum 
group $U_q(n)$, see \cite{ucr}. 
 
To see the mechanism of ultraviolet regularisation in the 
quantum field theoretical path integral, recall again that 
the functional analysis of representations of the 
quantum mechanical canonical $\x,\p$-commutation
 relations on wave
functions extends formally to representations on fields.
In the 
presence of a finite lower bound $\Delta x_0$ to localisability
the pointwise multiplication of fields is defined as
the maximally local multiplication.
The action functional and the path integral can be formulated
representation independently and the Feynman rules
follow straightforwardly.  
A crucial observation is that in the case III 
the above mentioned infinimum is indeed a minimum which is taken in $D$, i.e.
the maximally localised fields, formally written as $\vert x^{ml}\rangle$ 
obeying $\Delta x_{\vert x^{ml}\rangle} = \Delta x_0$ and
$\langle x^{ml}\vert \x \vert x^{ml}\rangle = x$, are normalisable \cite{go}.
It can be shown that as a consequence
the familiar distributions in the Feynman rules, 
such as $\delta(x-x^\prime) =\langle x^{ml}\vert x^{ml}\rangle$ and
$G(x,x^\prime)=\langle x^{ml}\vert 
1/(\p^2 +m^2)\vert x^{\prime ml}\rangle$ turn regular
functions. Thus their powers and products become well-defined,
thereby regularising the ultraviolet{ }\cite{ak-jmp-reg,go,km}.

{\flushleft{\large {\bf 5~~ 
Internal symmetries
} }}
\medskip\newline
The compatibility of this third short distance structure with internal symmetries 
has not yet been fully explored. However, there has already appeared
a possible new mechanism
for the emergence of  internal symmetry spaces: Intuitively,
the internal degrees of freedom
appear as the remnants of those external
degrees of freedom which one might have expected 
to have been `lost' through the ultraviolet cutoff:

Not in the domain $D$ of the external degrees of freedom,
but in $H$, the $\x_i$ do
have self-adjoint extensions. They and the eigenvectors
automatically
form representations of the unitary groups
 that map the now nonvanishing 
deficiency spaces onto another. The eigenvalues have orbits under
these groups, which is reminiscent of and possibly deeply related to
 the Kaluza Klein mechanism. 
A priori, however, the orbit attached to a point, as far as point can be resolved
on this geometry, is not in compactified
dimensions. Instead, these group orbits are within the "point's"
unresolvable small patch of fuzzyness of size $\Delta x_0$. 

In the generic case, the unitary extensions of the symmetric $\x$'s 
Cayley transforms
do not commute. Depending on the choice of short distance structure,
the group of unitaries which 
are functions of the $\x_i$ only, i.e. the 
group of local gauge transformations, is therefore 
no longer commutative, as in the Yang Mills
case. A detailed account is in progress.
 For preliminary results see \cite{nlis}.

\end{document}